\documentstyle[12pt]{article}

\title{The Higher Orders of the Theory of Strong Perturbations in
Quantum Mechanics and the Secularity Problem}

\author{Marco Frasca \\
Via E. Gattamelata,3 \\
00176 Roma (Italia) \\ }

\date{}

\begin{document}

\maketitle

\newpage

\begin{abstract}

We solve the higher order equations of the theory of the strong
perturbations
in quantum mechanics given in M.Frasca, Phys. Rev. A {\bf 45}, 43 (1992),
by
assuming that, at the leading order, the wave function goes adiabatically.
This is accomplished by deriving the unitary operator of adiabatic
evolution
for the leading order. In this way it is possible to show that at least
one of the causes of the problem of phase-mixing, whose effect is
the polynomial increase in time of the perturbation terms normally called
secularities, arises from the shifts
of the perturbation energy levels due to the unperturbed part of
the hamiltonian. An example is given for a two-level system that, anyway,
shows a secularity at second order also in the standard theory of small
perturbations. The theory is applied to the quantum analog of a classical
problem that can become
chaotic, a particle under the effect of two waves of
different amplitudes, frequencies and wave numbers.

\end{abstract}

\newpage

\section{Introduction}

In a recent paper [1] we showed that the theory of the strong
perturbations in
quantum mechanics [2] gives, at the leading order, an adiabatic behavior
for
the quantum system which is applied to. Beside, we showed, through the
interaction picture, that the validity condition $<H_0> \ll <V>$, being
the average taken on each eigenstate of the perturbation,
should be verified at any time. Actually, this is a set of conditions that
could give rise to secularities as we will show, being a secularity
a polynomial contribution in time in the perturbation term. The
secularities
are normally attributed to phase-mixing in the solution of the
Schr\"{o}dinger equation. Actually, a resummation can be
easily accomplished by comparing the result of the leading order of the
strong perturbation theory with the same in the interaction picture. The
only
effect, however, can be a harmless shift in the energy-levels of the
perturbation.

We can prove all that by deriving the unitary evolution operator for the
leading order, that is, the adiabatic evolution operator for the given
equation. In this way, the equations of higher orders in ref.[2] can be
integrated and the above cause of secularities pointed out. This problem
is exemplified through a two-level model that, however, still shows a
secularity at second order. However, it should be stressed that the
model here considered suffers similar problems from the standard small
perturbation theory too.

The full theory is applied to a quantum version of a well-known classical
model [3], that is a particle under the effect of two waves having
different amplitudes, frequencies and wave-numbers. We assume the
perturbation acting from the far past, being the particle free in that
limit. The model appears treatable by our method as is the one in ref.[4].

The paper is so structured. In sec.2 we derive the unitary evolution
operator and the higher order equations are solved. A possible origin of
secularities is given. In sec.3 we apply
the theory to the case of a two-level system in a constant perturbation
showing how secularities can be partially eliminated,
being one of the causes the one pointed out in sec.2.
In sec.4 we consider the quantized version of the classical model of the
particle in interaction with two waves.
In sec.5 some conclusions are drawn and the problem of the limitations of
the
theory is considered.

\section{Higher Order Terms and Secularities}

The general theory of strong pertrubations in quantum mechanics as
developed in ref.[2] considers a unitary evolution operator $U(t,t_0)$
such that
\begin{equation}
    VU(t,t_0) = i\hbar\frac{dU(t,t_0)}{dt} \label{eq:main}
\end{equation}
being $V$ the perturbation. The general hamiltonian of the system has
the form $H=H_0 + V$ with $H_0$ that may also depend on time,
so that a perturbation series could be derived as
\begin{eqnarray}
|\psi(t)> &=& U|\psi(t_0)> \nonumber \\
          &-& \frac{i}{\hbar}U\int_{t_0}^{t}dt'U^\dagger(t') H_0 U(t')
              |\psi(t_0)>
              \label{eq:series} \\
          &+& \left(-\frac{i}{\hbar}\right)^2
              U\int_{t_0}^{t}dt'U^\dagger(t') H_0 U(t')
              \int_{t_0}^{t'}dt''U^\dagger(t'') H_0 U(t'')
              |\psi(t_0)> \nonumber \\
          &+& \cdots \nonumber
\end{eqnarray}
or, introducing the time ordering operator $T$, as
\begin{eqnarray}
|\psi(t)> &=& U|\psi(t_0)> \nonumber \\
          &-& \frac{i}{\hbar}U\int_{t_0}^{t}dt'
              U^\dagger(t') H_0 U(t')|\psi(t_0)>
              \\ \label{eq:series_o}
          &+& \frac{1}{2}
              \left(-\frac{i}{\hbar}\right)^2U\int_{t_0}^{t}dt'
              \int_{t_0}^{t}dt''
              T(U^\dagger(t') H_0 U(t')
              U^\dagger(t'') H_0 U(t''))|\psi(t_0)> \nonumber \\
          &+& \cdots \nonumber
\end{eqnarray}
It is easy to see that the main problem is the determination of the
operator $U$ and then the computation of $U^\dagger H_0 U$ to go to
higher orders. This can be easily accomplished if we consider the main
results of ref.[1], that is, the leading order wave function,
$|\psi^{(0)}(t)> = U|\psi(t_0)>$, is just the following adiabatic one
\begin{equation}
    |\psi(t)> \sim \sum_n c_n e^{i\gamma_n}
                             e^{-\frac{i}{\hbar}\int_{t_0}^t dt'v_n(t')}
                             |n;t>
\end{equation}
being $c_n = <n;t_0|\psi(t_0)>$, $V|n; t> = v_n(t)|n; t>$ and
\begin{equation}
\gamma_n(t) = \int_{t_0}^t dt' <n; t'|i\frac{d}{dt'}|n; t'>.
\end{equation}
This should
be compared with the one obtained in the interaction picture that gives
\begin{equation}
    |\psi(t)> \sim \sum_n c_n e^{i\gamma_n}
                      e^{-\frac{i}{\hbar}\int_{t_0}^t dt'<n; t'|H_0|n,
t'>}
                      e^{-\frac{i}{\hbar}\int_{t_0}^t dt'v_n(t')} |n;t>
                     \label{eq:inter}
\end{equation}
where level shifts appear due to the unperturbed part of the hamiltonian.
We now show that these shifts could give rise to secularities for the
condition
$<n;t|H_0|n;t> \ll v_n(t)$.

Let us consider the evolution operator $U$ for an adiabatic evolution. It
is
a simple matter to see that we can write for eq.(\ref{eq:main})
\begin{equation}
    U(t, t_0) = \sum_n e^{i\gamma_n(t)}
                       e^{-\frac{i}{\hbar}\int_{t_0}^t dt'v_n(t')}
|n;t><n;t_0|
\end{equation}
then
\begin{eqnarray}
   U^\dagger H_0 U &=&  \sum_n <n;t|H_0|n;t>|n;t_0><n;t_0| +
                        \sum_{m,n,m\neq n} e^{i[\gamma_n(t)-\gamma_m(t)]}
\\
                    & & e^{-\frac{i}{\hbar}\int_{t_0}^t
dt'[v_n(t')-v_m(t')]}
                        <m;t|H_0|n;t>|m;t_0><n;t_0|. \nonumber
\end{eqnarray}
We fix our attention on the first term on the rhs of the above equation.
That
term, when substitued in eq.(\ref{eq:series}), at the first order gives
\begin{eqnarray}
    -\frac{i}{\hbar}U\int_{t_0}^t dt'
    \sum_n <n;t'|H_0|n;t'>|n;t_0><n;t_0|\psi(t_0)> &=& \\
    \sum_n \left(-\frac{i}{\hbar}\int_{t_0}^t dt'<n;t'|H_0|n;t'>\right)
           c_n e^{i\gamma_n}
           e^{-\frac{i}{\hbar}\int_{t_0}^t dt'v_n(t')}|n; t>         & &
           \nonumber
\end{eqnarray}
from which we recognize the second term of the series development of the
exponential of the level shifts in eq.(\ref{eq:inter}). This term could
give
rise to secularities in the perturbation series
if the shifts $<n; t|H_0|n; t>$ are time-independent as we
are going to show in the next section. It must be noticed that the above
term
is a direct application of the condition $<n;t|H_0|n;t> \ll v_n(t)$
and comes directly from the theory of strong perturbations.
So, as a rule, such terms should be simply resummed away.
This is accomplished without difficulty by computing the
level shifts and using eq.(\ref{eq:inter}) as leading order.
By comparing the level shifts with the energy levels of the
perturbation, or if the shifts are simply harmless,
we are able to realize if we can neglect such shifts. Otherwise,
we retain them and rewrite the evolution operator as
\begin{equation}
    U(t, t_0) = \sum_n e^{i\gamma_n(t)}
                       e^{-\frac{i}{\hbar}\int_{t_0}^t dt'<n; t'|H_0|n;
t'>}
                       e^{-\frac{i}{\hbar}\int_{t_0}^t dt'v_n(t')}
                       |n;t><n;t_0|
\end{equation}
redefining the full perturbation series. We will clarify the above
arguments
with the following example.

\section{Two-Level System with a Constant Perturbation}

We consider the hamiltonian
\begin{equation}
    H = H_0 + V = \left( \begin{array}{cc}
                            E_1 & 0 \\
                            0   & E_2
                         \end{array} \right) +
                  \left( \begin{array}{cc}
                            0   & V_{12} \\
                            V_{21} & 0
                         \end{array} \right)
\end{equation}
whose exact solution is well-known [5]. We apply to it the above results
to
exemplify the method.

The eigenstates of the perturbations are
\begin{equation}
   |v_1> = \frac{1}{\sqrt{2}}\left( \begin{array}{c}
                                            1 \\
                                            -\frac{V_{21}}{|V_{12}|}
                                    \end{array} \right),
   |v_2> = \frac{1}{\sqrt{2}}\left( \begin{array}{c}
                                            \frac{V_{12}}{|V_{12}|} \\
                                            1
                                    \end{array} \right)
\end{equation}
corresponding to the eigenvalue $-|V_{12}|$ and $|V_{12}|$ respectively.
Then, we have
\begin{equation}
    U(t) = e^{\frac{i}{\hbar}|V_{12}|t}|v_1><v_1| +
           e^{-\frac{i}{\hbar}|V_{12}|t}|v_2><v_2|
\end{equation}
and
\begin{eqnarray}
    U^\dagger(t)H_0U(t) &=& \frac{E_1 + E_2}{2}I + \label{eq:uh0u} \\
    & & \frac{E_1-E_2}{2|V_{12}|}
        \left(V_{12}e^{-2\frac{i}{\hbar}|V_{12}|t}|v_1><v_2|+
        V_{21}e^{2\frac{i}{\hbar}|V_{12}|t}|v_2><v_1|\right) \nonumber
\end{eqnarray}
being
\begin{equation}
    <v_1|H_0|v_1> = <v_2|H_0|v_2> = \frac{E_1 + E_2}{2}.
\end{equation}
So, the first term in the rhs of eq.(\ref{eq:uh0u}) is just the
contribution
from the level shifts that, by comparing with eq.(\ref{eq:inter}),
reduces simply to a harmless phase factor and can be systematically
neglected, the condition $|V_{12}| \gg \frac{E_1 + E_2}{2}$
to be compared with the exact solution of this problem.

Our method works till second order as does the standard small perturbation
theory as, at that order, a term increasing with time appears. So, we have
found a possible cause of secularities but the problem is still open.
However,
it should be stressed that secularities are a general problem also for the
standard small perturbation theory [6], the question is to understand
the origin of them. This does not mean at all that the method is unuseful
as
we are going to show in the next section.

\section{The Two-Wave Model}

We consider the hamiltonian
\begin{equation}
    H = \frac{p^2}{2m} + V_1 cos(k_1 x - \omega_1 t) +
                         V_2 cos(k_2 x - \omega_2 t)
\end{equation}
that can be easily quantized with the substitution
$p\rightarrow {\displaystyle -i\hbar\frac{\partial}{\partial x}}$.
This problem is analog
to the classical one in ref.[3] of a pendulum under the effect of an
oscillatory perturbation. We assume an adiabatic switching of the
perturbation from $t = -\infty$, being the particle free in
that limit, that is $\psi(x, -\infty) = \frac{1}{\sqrt{2\pi\hbar}}
e^{i\frac{p}{\hbar}x}$. This class of problems is tractable from the
point of view of secularities as already shown in ref.[4] and as we
are going to see in this case.

The leading order solution is then
\begin{eqnarray}
    \psi^{(0)}(x,t) &=&\sum_{m=-\infty}^{+\infty}
                       \sum_{n=-\infty}^{+\infty}
                       J_m\left(\frac{V_1}{\hbar\omega_1}\right)
                       J_n\left(\frac{V_2}{\hbar\omega_2}\right) \times
                       \nonumber \\
                    & &e^{-i(mk_1+nk_2)x}e^{i(m\omega_1+n\omega_2)t}
\times \\
                    & &\frac{1}{\sqrt{2\pi\hbar}}e^{i\frac{p}{\hbar}x}
                       \nonumber
\end{eqnarray}
being $J_m$ and $J_n$ the Bessel functions of order $m$ and $n$
respectively.
It is quite simple to derive
\begin{eqnarray}
    U^\dagger H_0 U \psi(x, -\infty) &=& \left(
    \frac{p^2}{2m} + \frac{k_1^2 V_1^2}{4m\hbar^2\omega_1^2} +
    \frac{k_2^2 V_2^2}{4m\hbar^2\omega_2^2}\right) \psi(x,-\infty) +
    \nonumber \\
    & &\sum_{\stackrel{m,r}{m\neq r}}\sum_{\stackrel{n,s}{n\neq s}}
    J_m\left(\frac{V_1}{\hbar\omega_1}\right)
    J_r\left(\frac{V_1}{\hbar\omega_1}\right)
    J_n\left(\frac{V_2}{\hbar\omega_2}\right)
    J_s\left(\frac{V_2}{\hbar\omega_2}\right) \times  \label{eq:J}\\
    & &\frac{(p-m\hbar k_1 -n\hbar k_2)^2}{2m} \times \nonumber \\
    & &e^{-i[(m-r)k_1+(n-s)k_2)]x}e^{i[(m-r)\omega_1+(n-s)\omega_2)]t}
    \psi(x,-\infty) \nonumber
\end{eqnarray}
having put
$H_0 = {\displaystyle -\frac{\hbar^2}{2m}\frac{\partial^2}{\partial
x^2}}$.
It easy to
see the part originating the secularities. This is the first term on the
rhs
that gives rise to a phase-factor due to the kinetic energy of the
particle
and the ponderomotive forces of the two waves. This term can be summed
away.
The sums in eq.(\ref{eq:J}) can be evaluated and this yields the
development
parameter of the perturbation series. We avoid this calculation here being
not the main point of the paper, we just point out that no secularities
appear at this order.

\section{Conclusions}

With the fundamental result of ref.[1], the theory of strong perturbations
in quantum mechanics is strictly linked with the adiabatic theorem. This
means that all the limitations of the adiabatic theorem should be ported
to
this theory. One problem may be due to a continuos spectrum of eigenvalues
for the perturbation. We have avoided to face this problem in sec.4,
although meaningful results was obtained.

However, the main question remains the secularities arising from
the perturbative solution of the Schr\"{o}dinger equation. We are not
assured that going to higher orders, secularities will not appear. We
would
like to stress again that this kind of problems arise normally in the
standard theory of small perturbations as could be seen in ref.[6] where
some methods are given to face the question.

Anyway, we showed with a lot of examples that the theory is indeed
useful and a wide possibility to explore new solutions of the
Schr\"{o}dinger equation is surely open. This in turn means that new
quantum behaviors should be considered, that is, the class of quantum
systems
that we define strongly perturbed.

\newpage

[1] M.Frasca, ``{\sl The Leading Order of the Theory of Strong
Perturbations in Quantum Mechanics}'', quant-ph/9507007 (1995);
M.Frasca, submitted to Nuovo Cimento (1995);
see also A.Joye, Ann. Inst. Henri Poincare', {\bf 63}, 231 (1995)

[2] M.Frasca, Phys. Rev. A {\bf 45}, 43 (1992);
Phys. Rev. A {\bf 47}, 2364 (1993)

[3] R.Z.Sagdeev, D.A.Usikov, G.M.Zaslavsky,
{\sl Nonlinear Physics (From the Pendulum to Turbulence and Chaos)},
Harwood Academic Publishers, Philadelphia, Penn., 1992

[4] M.Frasca, Nuovo Cimento B, {\bf 107}, 845 (1992);
Nuovo Cimento B, {\bf 109}, 1227 (1994); Phys. Rev. E, {\bf 53}, 1236
(1996)

[5] J.J.Sakurai, {\sl Modern Quantum Mechanics}, Benjamin/Cummings
Publishing Company, Reading, Mass., 1985

[6] J.Kevorkian, J.D.Cole, {\sl Perturbation Methods in Applied
Mathematics}, Springer-Verlag, New York, N.Y., 1985

\end{document}